\documentclass{article}

\newcommand{\SSM}{Supersymmetric Standard Model}

\newcommand{\bt}{\begin{tabular}{c}}
\newcommand{\et}{\end{tabular}}

\newcommand{\eb}{\ee\be } 
\newcommand{\ebp}{\rt.\ee\be\lt.} 
\newcommand{\bmat}{\lt ( \begin{array} }
\newcommand{\emat}{  \end{array} \rt )}

\newcommand{\oQ}{{\ov Q}}

\newcommand{\ovD}{{\ov D}}

\newcommand{\cd}{{\cdot}}

\newcommand{\oM}{{\ov M}}

\newcommand{\ot}{{\ov t}}

\newcommand{\oq}{{\ov \q}}

\newcommand{\ED}{\end{document}}

\newcommand{\oC}{{\ov C}}

\newcommand{\A}{{\ov A}}

\renewcommand{\a}{\alpha}	
\renewcommand{\b}{\beta}
\newcommand{\g}{\gamma}
\renewcommand{\d}{\delta}
\newcommand{\e}{\epsilon}

\newcommand{\z}{\zeta}
\newcommand{\h}{\eta}
\newcommand{\q}{\theta}

\newcommand{\y}{\psi}
\newcommand{\w}{\omega}
\newcommand{\G}{\Gamma}
\newcommand{\D}{\Delta}
	
\renewcommand{\L}{\Lambda}

\newcommand{\la}{\label}
\newcommand{\ci}{\cite}

\newcommand{\ds}{\documentstyle}	
\newcommand{\fr}{\frac}

\newcommand{\pa}{\partial}
\newcommand{\ov}{\overline}
\newcommand{\be}{\begin{equation}}
\newcommand{\ee}{\end{equation}}
\newcommand{\ba}{\begin{array}} 
\newcommand{\ea}{\end{array}}
\newcommand{\bea}{\begin{eqnarray}}
\newcommand{\eea}{\end{eqnarray}}
\newcommand{\ra}{\rightarrow}
\newcommand{\Ra}{\Rightarrow}

\newcommand{\lt}{\left}
\newcommand{\rt}{\right}

\newcommand{\ben}{\begin{enumerate}}
\newcommand{\een}{\end{enumerate}}
\newcommand{\bitem}{\begin{itemize}}
\newcommand{\eitem}{\end{itemize}}

\newcommand{\articletitle}{Some Properties of  Chiral Dotted  Spinor Superfields}

\begin{document}
\makeatletter	   
\renewcommand{\ps@plain}{%
\renewcommand{\@evenhead}{\@oddhead}
\renewcommand{\@evenfoot}{\@oddfoot}}
\makeatother    
\title{  \articletitle }
\author{ John A.  Dixon\\ Dixon Law Firm \\1020 Canadian Centre\\
833 - 4th Ave. S. W. \\ Calgary, Alberta \\ Canada T2P 3T5 }
\maketitle
\pagestyle{plain}

\abstract {\normalsize
 Chiral superfields with multiple dotted Lorentz spinor indices (`dotspinors') are important in the analysis of supersymmetry breaking through the mechanisms of Cybersusy.  This paper describes the  actions for massive dotspinors coupled to  supersymmetric gauge theory and to chiral matter.  It analyzes  the free equations of motion and mass spectra for the case of unbroken supersymmetry. The general form of the Cybersusy algebra for dotsupers with multiple indices is also discussed briefly.}

\large

\section{Introduction}

\subsection{Dotspinors}

In many ways the chiral dotted spinor superfields are natural generalizations of the well known chiral scalar superfield. In this paper we shall sometimes use the word `dotspinors' to mean   `chiral dotted spinor superfields'.  

These dotspinors are of two kinds: dotsupers and dotpseudos.  
The term `dotsupers' will be used to designate dotspinors which do not contain Zinn-Justin fields.  The term `dotpseudos' will be used to designate dotspinors which do contain Zinn-Justin fields.  

There are two series of dotspinors, and this is true for both dotsupers and dotpseudos.   The first series  of dotspinors contains the bosonic dotspinors ${\hat A}_{(\dot \a_1 \cdots \dot \a_{2n})}$ with integer spin $n=0,1,2\cdots$. The second series of dotspinors contains the fermionic dotspinors ${\hat \w}_{(\dot \a_1 \cdots \dot \a_{2n+1})}$ with half integer spin 
$\fr{2n+1}{2}=\fr{1}{2},\fr{3}{2},\fr{5}{2},\cdots$. 

The bosonic chiral superfield  ${\hat A}_{(\dot \a_1 \cdots \dot \a_{2n})}$ is symmetric under any permutation of its 2n indices $(\dot \a_1 \cdots \dot \a_{2n})$. It  satisfies the chiral constraint equation:
\be
\ov D_{\dot \b} {\hat A}_{(\dot \a_1 \cdots \dot \a_{2n})}=0, n= 0, 1, \cdots
\la{evenspin}
\ee
The well-known chiral scalar superfield ${\hat A}$ is the n=0 version of (\ref{evenspin}). It has no dotted index at all. Chiral scalar superfields ${\hat A}^i$ are used to make the matter fields, such as the quarks and leptons, in models like the \SSM \ (`SSM').  The fermionic chiral superfields  ${\hat \w}_{(\dot \a_1 \cdots \dot \a_{2n+1})}$ are also symmetric under any permutation of their 2n+1 indices $(\dot \a_1 \cdots \dot \a_{2n+1})$.  They also satisfy the chiral constraint equations:
\be
\ov D_{\dot \b} {\hat \w}_{(\dot \a_1 \cdots \dot \a_{2n+1})}=0, n= 0, 1, \cdots
\ee

\subsection{Components for dotsupers}

Because the  dotspinors are chiral, they have a simple expansion in terms of components:
\[
 {\hat A}_{\dot \a_1 \cdots \dot \a_{2n}}(x)
=
 A_{\dot \a_1 \cdots \dot \a_{2n}}(y)
\]\be
+ \q^{\a}  \y_{\a, \dot \a_1 \cdots \dot \a_{2n}}(y) 
+ \fr{1}{2} \q \cd \q F_{\dot \a_1 \cdots \dot \a_{2n}}(x)
\la{evenspinexpansion}
\ee

\[
 {\hat \w}_{\dot \a_1 \cdots \dot \a_{2n+1}}(x)
=
 \w_{\dot \a_1 \cdots \dot \a_{2n+1}}(y)
\]\be
+ \q^{\a}  W_{\a, \dot \a_1 \cdots \dot \a_{2n+1}}(y) 
+ \fr{1}{2} \q \cd \q \L_{\dot \a_1 \cdots \dot \a_{2n+1}}(x)
\la{oddspinexpansion}
\ee
The superderivatives are: 
$
 D_{\a} = \fr{\pa}{\pa \q^{\a}} +  \fr{1}{2} \pa_{\a \dot \b} \oq^{\dot \b} $ and $  \ov D_{\dot \a} = \fr{\pa}{\pa \oq^{\dot \a}} +  \fr{1}{2} \pa_{\b \dot \a} \q^{\b}
$.
Here  the chiral translated spacetime variable is $y_{\a \dot \b}  = x_{\a \dot \b} + \fr{1}{2} \q_{\a } \ov \q_{\dot\b} 
$. It satisfies 
\be
 \ov D_{\dot \a} y_{\a \dot \b}=0
\ee
This is a simple extension of the results in such standard works as \ci{WB}
\ci{west}\ci{superspace}\ci{ferrara}.

\subsection{A little about Cybersusy}

The chiral dotted spinor superfields with $n\neq 0$  have largely been ignored during the development of supersymmetric quantum field theory.  This is understandable, because these superfields generally describe higher spin supermultiplets without any gauge invariance.  However, these dotsuper multiplets (for all $n$), play an important role in  the BRS cohomology analysis of composite operators in models like the supersymmetric standard model.  In that context, they are crucial for  the analysis of supersymmetry breaking through the mechanisms of Cybersusy\footnote{This series of five papers needs to be revised along the lines described briefly in \ci{susy09}.  Essentially the problem is that the SSM  breaks only the right half of the supersymmetry, and this was not noticed in these five papers. Both halves probably need to be broken to generate a breaking which is consistent with experiment.  The resolution of this issue probably requires a model which changes the SSM somehow. It is not presently known whether such a model exists. A little more discussion of this problem can be found in section \ref{guesses} below.} \ci{cybersusyI}\ci{cybersusyII}\ci{cybersusyIII}\ci{cybersusyIV}\ci{cybersusyV}. These dotspinors may also be relevant to the description of the higher spin supersymmetry multiplets of the four dimensional superstring.

Cybersusy is based on an observation, two ideas and a hope:
\ben
\item
Firstly one observes that many  dotpseudos arise in the  BRS cohomology of a theory like the SSM, when the auxiliaries have been integrated.  These dotpseudos bear a close resemblance to the observed particles of our world, and generally they mix with 
the observed particles as soon as gauge invariance is spontaneously broken by the VEV of a Higgs/Goldstone type multiplet. This mixing gives rise to an algebra of the form:
\be
 \d_{\rm BRS} = 
\d_{\rm SUSY} +  \d_{\rm GSB} 
\ee
 linking the dotspseudos with each other. 

\item
Then the first idea is to promote these dotpseudos to  dotsupers and write down an effective theory based on the algebra 
and properties of the dotpseudos.  The new algebra is
\be
 \d_{\rm CYBERSUSY} = 
\d_{\rm SUSY} +  \d_{\rm MIX} 
\ee

\item
The second idea  is that it is necessary to include a dotspinor mass term  ${\cal A}_{\rm MASS}$ in the effective action, because otherwise one has a massless supermultiplet in the theory.
Supersymmetry breaking arises from this term ${\cal A}_{\rm MASS}$ in the effective action.   When  $ \d_{\rm MIX} $ acts on the effective action,  there is no local polynomial solution ${\cal A}_{\rm Counterterm}$ for the equation:
\be
   \d_{\rm MIX}  {\cal A}_{\rm MASS}
+
\d_{\rm SUSY} {\cal A}_{\rm Counterterm}
=0
\ee
So the term 
\be
{\cal A}_{\rm Anomaly} =  \d_{\rm MIX}  {\cal A}_{\rm MASS}
\ee
acts like a supersymmetry anomaly in the theory. The breaking follows from the presence of this supersymmetry anomaly, which behaves  rather like explicit breaking of supersymmetry.
\item
The hope is that no unsolvable problem arises in this scheme.  Somehow the theory may make sense in spite of the presence of this breaking.  One concern is whether the anomalies spoil unitarity somehow.  This needs investigation, of course.

\een

\subsection{Higher spin without gauge invariance?}

Cybersusy and dotsupers suggest that gauge invariance is not the only way to deal with vector or higher spin particles.  There are `cluster propagators' that are relevant to such particles, and they give rise to  another way to treat higher spin, without invoking gauge invariance.    The cluster propagators mix up several fermions or they mix scalars and vectors in complicated ways. Moreover cybersusy appears to generate a way of describing broken supersymmetry using these clusters.  Examples can be found in \ci{cybersusyIV}.  It is expected that such cluster propagators exist for higher spins also.

\subsection{A little more about dotsupers and dotpseudos and composite operators}

As mentioned above, the following kinds of chiral dotted spinor superfields (dotspinors) arise:
\ben
\item
Elementary chiral dotted spinor superfields, which are made from elementary components.  These will be called  simply `dotsupers'.  Sometimes we will call these `elementary dotsupers' when it is important to distinguish them from the composite dotsupers.  These were already used in the cybersusy papers to write down effective actions.  Here we shall show how to write down these actions coupled to Yang Mills type  supersymmetric gauge theory. 
\item
Composite chiral dotted spinor superfields made from other superfields using the chiral derivatives $\ov D_{\dot \a}$.  
 These will be called `composite dotsupers'. These will be introduced in this paper in section \ref{compdotsupers} .  It is natural to write these down and to then couple them to elementary dotsupers.  This results in a coupling between dotsupers and elementary chiral matter.
\item
Composite chiral dotted spinor superfields made from the components of other superfields and also  the Zinn Justin sources and their variations.  These will be called `dotpseudos'. 

In the Cybersusy papers we wrote down elementary dotsupers corresponding to these dotpseudos, and then wrote down actions for the dotsupers, using the cybersusy algebra that arose for the dotpseudos from the BRS transformations after gauge symmetry breaking. 

As we shall discuss in section  \ref{compdotsupers}, it does not make sense to couple dotpseudos to dotspinors.
\een

\subsection{Contents of this Paper}

This paper is mostly concerned with an easy topic:  how to construct couplings between dotsupers and ordinary chiral matter on the one hand, and Yang-Mills type vector gauge supersymmetry on the other hand. 

These Yang-Mills couplings lead naturally to the free equations of motion of dotspinors, and also they naturally accomodate the concept of mass for dotspinors.

One interesting feature we will find is that  the actions for dotsupers have an amusing way of defining a propagator with a unique mass for each component of the superfield, in spite of the fact that there are multiple factors of $\D= \fr{1}{2} \pa_{\a \dot \b} \pa^{\a \dot \b}$  in the propagator. This is undoubtedly important for cybersusy, since it allows the theory to develop a reasonable spectrum before and after supersymmetry breaking. In this context, a conjecture arises relating to the distribution of solutions for certain polynomial equations. 

The connection between dotspinors and chiral matter, and the relation of this to Cybersusy is still quite obscure.  So all we can do here is look briefly at the coupling for simple cases.

An important problem remains unsolved.  How can one find a model like the SSM which gives rise to the cybersusy algebra for both left and right sectors?

This paper also prepares the ground for the Cybersusy analysis of supersymmetry breaking for various higher spin supermultiplets,  including  the baryons, the mesonic hadrons,  the gauge bosons and the Higgs particles.  However, in this introductory paper we shall largely assume that  gauge symmetry and supersymmetry are unbroken. 
\

\section{ Dotsupers Coupled to Supersymmetric Yang Mills Gauge Theory}

\la{sourcechapter} 

Many of the dotpseudos  that we find in Cybersusy  transform under U(1) gauge transformations. So far in the papers  \ci{cybersusyI}\ci{cybersusyII}\ci{cybersusyIII}\ci{cybersusyIV}\ci{cybersusyV}\ci{susy09},  only free massive effective actions have been written down. One takes the algebra from the dotpseudos and writes down actions for dotsupers with the same algebra to generate these free massive effective actions.

How can one extend this to an interacting effective action with supersymmetry breaking? 

As a first step in this direction, we want to couple dotsupers to a supersymmetric gauge theory.  

 Actually we can construct invariant actions for dotsupers coupled to any compact gauge group, so we will present the more general result here.  We start with the well-known result for  chiral scalar superfields ${\hat A}$ to establish the notation.
These results for ${\hat A}$  can be found  in standard works such as \ci{WB}
\ci{west}\ci{superspace}\ci{ferrara}.

\subsection{ Chiral Scalar Superfields ${\hat A}$  Coupled to Gauge Theory}
\la{scalargaugeaction}
Suppose that the gauge transformation\ci{WB}\ci{west}\ci{superspace}\ci{ferrara} of ${\hat A}$  has the form:
\be
{\hat A}  \Ra 
e^{i{g}  {\hat S}} {\hat A}
\la{dfdsffsfsdsclar}
\ee
This equation is short for
\be
{\hat A}^k \Ra 
(e^{i{g}  {\hat S}^a t^a})^k_{\;\;l} 
{\hat A}^l
\la{dfdsffsfsdindicessclar}
\ee
where we suppose that $t^{ai}_{\;j}$
are hermitian matrices that form a representation of some compact group, with commutator 
\be 
t^{ai}_{\;\;j} 
t^{bj}_{\;\;k} 
-
t^{bi}_{\;\;j} 
t^{aj}_{\;\;k} 
=
i f^{abc} t^{cj}_{\;\;k} 
\ee
We also suppose that $ {\hat S}^a$  
and  ${\hat A}^i$ are Grassmann even chiral scalar   superfields:
\be
\ovD_{\dot \b} 
{\hat S}^a
=\ovD_{\dot \b} {\hat A}^i =
0
\ee
So this means that ${\hat S}^a$ is a vector in the adjoint representation of the gauge group, and that $  {\hat A}^i $ is a chiral superfield in the  representation of that group characterized by $t^{ai}_{\;\;j}$.

The factor $e^{i{g}  {\hat S}} $ looks a bit like a  unitary representation.  But it is not.
${\hat S}$ is a chiral superfield, which means that it is complex.  Even if we restrict  ${\hat S}\ra S \equiv {\hat S}_{|_{\q = \ov \q =0}}$ to its  scalar part, so that there are no  complications resulting from the superspace components $\q_{\a}$ and $ \ov \q_{\dot \a} $, then this is still not  a unitary representation.  That would require that  $ S^a  $ be  a real vector  $S^a= \ov S^a$ contracted with the hermitian matrices $t^{ai}_{\;j}$.  But that is  inconsistent with the  intrinsically complex nature of ${\hat S}^a$ and ${S}^a$. 

In general for the non-Abelian case we can write the gauge transformation in the form:
\be
e^{-{g} {\hat V}}  
\la{noihere}
\Ra
e^{  i {g}  {\hat {\ov S}}  }
e^{  -{g} {\hat V}   }
e^{- i {g}  {\hat {S}} }
\ee
Here V is taken to be  the matrix
\be
V \equiv V^a t^{ai}_{\;\;j}
\ee
where ${\hat V}$ is a real superfield:
\be
{\hat V}^a = ({\hat V}^a)^* \equiv {\ov {\hat V}}^a \equiv {\hat {\ov V}}^a 
\ee

Note the absence of a factor $i$ in the numerator of the matrix $e^{-{g} {\hat V}} $  in (\ref{noihere}). The matrix $e^{-{g} {\hat V}} $ is not unitary even though  
$ ({\hat V}^a)_{|_{ \q = \ov \q =0} } = V^a$ is real and the matrices $t^{ai}_{\;\;j}$ are hermitian. However the matrix 
$e^{-{g} {\hat V}} $ is hermitian (in a sense, if one does not worry too much about the meaning of $\q$ and $\ov \q$). 
 The inverse matrix is
\be
e^{{g} {\hat V}}  
\Ra
e^{ i {g}  {\hat {S}} }
e^{  {g} {\hat V}   }
e^{ - i {g}  {\hat {\ov S}}  }
\ee
Then we have:
\be
e^{-{g} \hat V}  {\hat A}  
\Ra e^{  i {g}  {\hat {\ov S}}  }
e^{  -{g} {\hat V}   }
e^{- i {g}  {\hat {S}} }
e^{i{g} {\hat S}} {\hat A}
= e^{ i {g} {\hat {\ov S}}  }
\lt ( e^{ -{g} {\hat V}   }
  {\hat A} \rt )
\ee

Now the complex conjugate of 
(\ref{dfdsffsfsdsclar}) is

\be
{\hat {\ov A}}_{k } \Ra 
(e^{- i{g}  {\hat {\ov S}}^a (t^{a})^*})_k^{\;\;l} {\hat {\ov A}}_{l }  \equiv
(e^{- i{g}  {\hat {\ov S}}^a t^{a}})^l_{\;\;k} {\hat {\ov A}}_{l } \equiv
{\hat {\ov A}}_{l }  (e^{- i{g}  {\hat {\ov S}}^a t^{a}})^l_{\;\;k} 
\la{dfdsffsfsdindicescc}
\ee
which we abbreviate to:
\be
{\hat {\ov A}}^T
\Ra 
 {\hat {\ov A}}^T
 e^{- i{g}  {\hat {\ov S}}} 
\ee
It follows that:
\be{\hat {\ov A}}^T
e^{-{g}\hat V}  
\Ra 
 {\hat {\ov A}}^T
 e^{- i{g}  {\hat {\ov S}}} 
  e^{  i {g}  {\hat {\ov S}}  }
e^{  -{g} {\hat V}   }
e^{- i {g}  {\hat {S}} }
\eb
= {\hat {\ov A}}^T
e^{-{g}\hat V}  
e^{- i {g} {\hat S}  }
\ee
and so
 we see that the well known form
\be
{\cal A} =
-\int d^4 x \; d^4 \q \; 
\lt \{
{\hat {\ov A}}^T
e^{-{g}\hat V}  
 {\hat A}
\rt \}
\la{eryhrthrt}
\ee
is invariant under the gauge transformations and that it is also invariant under supersymmetry. The component form of the above, in the Wess-Zumino gauge, for example, can be obtained by projection in the usual way.

\subsection{ Chiral Dotted Spinor Superfields ${\hat \w}_{\dot \a}$   Coupled to Gauge Theory}

\la{dotsupergaugeaction}

Now we discuss the generalizations of the scalar superfield, starting with the chiral dotted spinor superfield  ${\hat \w}_{\dot \a}$ .

Suppose that the gauge transformation of ${\hat \w}_{\dot \a}$  has the form:
\be
{\hat \w}_{\dot \a} \Ra 
e^{i{g}  {\hat S}} {\hat \w}_{\dot \a}
\la{dfdsffsfsd}
\ee
This equation is short for
\be
 {\hat \w}^k_{\dot \a} \Ra 
\lt (e^{i{g}  {\hat S}^a t^a} \rt )^k_{\;\;l} {\hat \w}^l_{\dot \a}
\la{dfdsffsfsdindices}
\ee

We also suppose that $ {\hat S}^a$ is a Grassmann even chiral scalar superfield 
and  ${\hat \w}^i_{\dot \a}$ is a Grassmann odd chiral dotted spinor superfield:
\be
\ovD_{\dot \b} 
{\hat S}^a
=\ovD_{\dot \b} {\hat \w}^i_{\dot \a} =
0
\ee
The dotsuper ${\hat \w}^i_{\dot \a} $ is a vector in the representation space characterized by the hermitian matrices 
$ t^{ai}_{\;\;j}$ just like the scalar superfield  was in 
(\ref{dfdsffsfsdsclar}).

Again, as for the chiral scalar superfield, we have:
\be
e^{-{g} \hat V}  {\hat \w}_{\dot \a}  
\Ra e^{  i {g}  {\hat {\ov S}}  }
e^{  -{g} {\hat V}   }
e^{- i {g}  {\hat {S}} }
e^{i{g} {\hat S}} {\hat \w}_{\dot \a}
= e^{ i {g} {\hat {\ov S}}  }
\lt ( e^{ -{g} {\hat V}   }
  {\hat \w}_{\dot \a} \rt )
\ee
But now we need more indices, so we take 

\be
\ov D_{\dot \b} e^{{g} \hat V} D_{\a} e^{-{g} \hat V}  {\hat \w}_{\dot \a}  
\Ra 
\ov D_{\dot \b} e^{ i {g}  {\hat {S}} }
 e^{{g} \hat V} e^{ - i {g}  {\hat {\ov S}}  }
 D_{\a} 
e^{  i {g}  {\hat {\ov S}}  }
e^{  -{g} {\hat V}   }
e^{- i {g}  {\hat {S}} }
e^{i{g} {\hat S}} {\hat \w}_{\dot \a}
\eb
= e^{ i {g} {\hat { S}}  }
\lt (\ov D_{\dot \b} e^{{g} \hat V} D_{\a} e^{-{g} \hat V}  {\hat \w}_{\dot \a}   \rt )
\ee

So we see that it is useful to define a covariant double chiral derivative of the form 
\be
{\cal D}_{\a \dot \b} \equiv
\ov D_{\dot \b} e^{{g} \hat V} D_{\a} e^{-{g} \hat V}
\ee
and we note that it tranforms in the simple manner:
\be
{\cal D}_{\a \dot \b} 
\Ra
 e^{ i {g} {\hat { S}}  }
{\cal D}_{\a \dot \b} 
 e^{- i {g} {\hat { \ov S}}  }
\ee

The complex conjugate of 
(\ref{dfdsffsfsdindices}) is 
\be
{\hat {\ov \w}}_{k \a} \Ra 
(e^{- i{g}  {\hat {\ov S}}^a (t^{a})^*})_k^{\;\;l} {\hat {\ov \w}}_{l  \a} \equiv
(e^{- i{g}  {\hat {\ov S}}^a t^{a}})^l_{\;\;k} {\hat {\ov \w}}_{l  \a}\equiv
{\hat {\ov \w}}_{l  \a} (e^{- i{g}  {\hat {\ov S}}^a t^{a}})^l_{\;\;k} 
\la{dfdsffsfsdindicescc2}
\ee
which we abbreviate to:
\be
\lt ( {\hat {\ov \w}}_{\a} \rt )^T
\Ra 
\lt ( {\hat {\ov \w}}_{\a} \rt )^T e^{- i{g}  {\hat {\ov S}}} 
\ee
It follows that:
\be
\lt ( {\hat {\ov \w}}_{\a} \rt )^T
e^{-{g}\hat V}  
\Ra 
\lt ( {\hat {\ov \w}}_{\a} \rt )^T e^{- i{g}  {\hat {\ov S}}} 
  e^{  i {g}  {\hat {\ov S}}  }
e^{  -{g} {\hat V}   }
e^{- i {g}  {\hat {S}} }
\eb
= \lt ( {\hat {\ov \w}}_{\a} \rt )^T
e^{-{g}\hat V}  
e^{- i {g} {\hat S}  }
\ee
So we see that
\be
{\cal A} =
\int d^4 x \; d^4 \q \; 
\lt \{
\lt ( {\hat {\ov \w}}_{\a} \rt )^T
e^{-{g}\hat V}  
\rt \}
\lt \{
 {\ov D}_{\dot \b}
e^{{g}\hat V} 
D^{\a} e^{-{g}\hat V}  {\hat \w}^{\dot \b}
\rt \}
\eb
=
\int d^4 x \; d^4 \q \; 
\lt \{
{\hat {\ov \w}}_{\a}^T
e^{-{g}\hat V}  
 {\ov D}_{\dot \b}
e^{{g}\hat V} 
D^{\a} e^{-{g}\hat V}  {\hat \w}^{\dot \b}
\rt \}
\eb
\equiv
- \int d^4 x \; d^4 \q \; 
\lt \{
{\hat {\ov \w}}^{\a T}
e^{-{g}\hat V}  
{\cal D}_{\a \dot \b} \; {\hat \w}^{\dot \b}
\rt \}
\ee
is invariant under the gauge transformations.  It is also invariant under supersymmetry.
We contracted the indices to form a Lorentz invariant here.

So this is the kinetic action for the dotsuper with one dotted index, coupled to a supersymmetric gauge theory.

\subsection{ Dotsupers with any number of indices, coupled to Gauge Theory}

It is easy to perform the similar procedure for doubledotsupers.  Suppose that the gauge transformation is:
\be
{\hat A}_{\dot \a\dot \b} \Ra 
e^{i{g}  {\hat S}} 
{\hat A}_{\dot \a\dot \b}
\ee
The action is
\be
{\cal A} = 
\int d^4 x \; d^4 \q \; 
\lt (
 ({\hat \A}^{ \g \d })^{ T}  e^{-{g} \hat V}  
{\cal D}_{\g \dot \a}  
{\cal D}_{\d \dot \b}  
  {\hat A}^{\dot \a\dot \b} 
 \rt )
\ee
This is easy to generalize to an arbitrary dotsuper with any number of indices.
The symmetry ${\hat A}^{\dot \a\dot \b} ={\hat A}^{\dot \b\dot \a} $ means that the choice of distribution of  the indices for the operators 
${\cal D}_{\g \dot \a}  {\cal D}_{\d \dot \b}  
$ ( which could be, say, ${\cal D}_{\d \dot \a}  {\cal D}_{\g \dot \b}  
$)  does not make any difference to the result.

The generalization to dotsupers with any number of indices is obvious.  For bosonic dotsupers we have
\be
{\cal A} =
\int d^4 x \; d^4 \q \; 
\lt (
 ({\hat \A}^{ a_1 \cdots \a_{2n}  })^{ T}  e^{-{g} \hat V}  
{\cal D}_{\a_1 \dot \b_1}  
\cdots
{\cal D}_{\a_{2n} \dot \b_{2n}}  
  {\hat A}^{\dot \b_1 \cdots \dot \b_{2n}} 
 \rt )
\ee
and for fermionic dotsupers we have
\be
{\cal A} =
\int d^4 x \; d^4 \q \; 
\lt (
 ({\hat {\ov \w}}^{ a_1 \cdots \a_{2n+1}  })^{ T}  e^{-{g} \hat V}  
{\cal D}_{\a_1 \dot \b_1}  
\cdots
{\cal D}_{\a_{2n+1} \dot \b_{2n+1}}  
  {\hat \w}^{\dot \b_1 \cdots \dot \b_{2n+1}} 
 \rt )
\ee

\section{Actions for dotsupers with Masses and Gauge Interactions}

Here we will add mass terms by doubling the number of dotsupers with a given spin, and choosing them so that there is a gauge invariant mass term available.

\subsection{ Action and Mass term for two Scalar Superfields ${\hat A}_{{\rm L} },{\hat A}_{{\rm R}}$ }

Suppose that we have two copies of the superfield described in subsection \ref{scalargaugeaction},  except that they have different quantum numbers: 
\be
{\hat A}_{{\rm L} } \Ra 
e^{i{g}  {\hat S}_{\rm L}} {\hat A}_{{\rm L}}
\la{scalarleft}\ee
\be
{\hat A}_{{\rm R} } \Ra 
e^{i{g}  {\hat S}_{\rm R}} {\hat A}_{{\rm R}}
\la{scalarright}\ee
Here we define 
\be
{\hat S}_{\rm L}= ({\hat S}^a t_{\rm L}^a)^k_{\;\;l} 
\ee
\be
{\hat S}_{\rm R}= ({\hat S}^a t_{\rm R}^a)^k_{\;\;l} 
\ee
and we will need:
\be
V_{\rm L} \equiv V^a t^{ai}_{{\rm L}\;\;j}
\ee\be
V_{\rm R} \equiv V^a t^{ai}_{{\rm R}\;\;j}
\ee
where the two representations are in general different.  For example if the group  is simply U(1), then we would take 
something like this:
\be
{\hat S}_{\rm L}= - q {\hat S} 
\ee
\be
{\hat S}_{\rm R}= + q{\hat S}
\ee
where $q$ is simply a number, so that the two could cancel in a mass term as set out below, but if the group is SU(3) and the matrices are the $3 \times 3$ representation of SU(3), we would want one of the representations to be a $3$ and the other a $\ov 3$ so that the mass term could be invariant using the invariant SU(3) tensor  $\d^i_j$.  

Then we take the action
\be
{\cal A}_{\rm L} =
-\int d^4 x \; d^4 \q \; 
\lt \{
{\hat \A}_{\rm L}^T
e^{-{g}{\hat V}_{\rm L} }  
 {\hat A}_{{\rm L} }
\rt \}
\la{scalarleftaction}
\ee
\be
{\cal A}_{\rm R} =
-\int d^4 x \; d^4 \q \; 
\lt \{
{\hat \A}_{\rm R}^T 
e^{-{g}{\hat V}_{\rm R} }  
 {\hat A}_{{\rm R} }
\rt \}
\la{scalarrightaction}
\ee

\be
{\cal A}_{\rm Mass} =
-\int d^4 x \; d^4 \q \; 
m \lt \{
{\hat A}_{\rm L}^T
M 
 {\hat A}_{{\rm R} }
\rt \} + *
\la{scalarmassterm}
\ee
Here $M$ is  an invariant numerical tensor of the gauge group in the following sense:
\be
t^{aj}_{{\rm L}\;\;i} M_{jk} 
+M_{ij} t^{aj}_{{\rm R}\;\;k}
=0
\ee
This action is then invariant under supersymmetry and  gauge symmetry. For a group like SU(3), if ${\hat A}_{\rm L}^{i }$ is in the triplet $3$ representation, it would make more sense to put the indices   down for the   right parts ${\hat A}_{i \rm R}$, which would need to be in the $\ov 3$ representation to form a mass term.

\subsection{ Action and Mass term for two Chiral Dotted Spinor Superfields ${\hat \w}_{{\rm L}\; \dot \a},{\hat \w}_{{\rm R}\; \dot \a}$  with one spinor index  }

\la{singledotsuperaction}

Now suppose we have two copies 
of the superfield described in subsection 
\ref{dotsupergaugeaction}, except that they have different quantum numbers: 
\be
{\hat \w}_{{\rm L}\; \dot \a} \Ra 
e^{i{g}  {\hat S}_{\rm L}} {\hat \w}_{{\rm L}\; \dot \a}
\la{dfdsffsfsdleft}\ee
\be
{\hat \w}_{{\rm R}\; \dot \a} \Ra 
e^{ i {g}  {\hat S}_{\rm R}} {\hat \w}_{{\rm R}\; \dot \a}
\la{dfdsffsfsdright}\ee

We use ${\hat S}_{\rm L}$, ${\hat S}_{\rm R}$, 
$V_{\rm L} $, and 
$V_{\rm R}$ just as for the previous subsection.

We can define the combinations
\be
{\cal D}_{\rm L}^{\dot \a \a}
\equiv  {\ov D}^{\dot \a}
 e^{{g} {\hat V_{\rm L}}} 
 D^{\a}
 e^{-{g} {\hat V_{\rm L}}}
\ee
and
\be
{\cal D}_{\rm R}^{\dot \a \a}
\equiv  {\ov D}^{\dot \a}
 e^{{g} {\hat V_{\rm R}}} 
 D^{\a}
 e^{-{g} {\hat V_{\rm R}}}
\ee
As we saw above, these transform under gauge transformations as
\be
{\cal D}_{\rm L}^{\dot \a \a}
\Ra
 e^{   i {g} {\hat S}_{\rm L} }
{\cal D}_{\rm L}^{\dot \a \a}
 e^{ -  i {g} {\hat {\ov S}}_{\rm L} }
\ee
and
\be
{\cal D}_{\rm R}^{\dot \a \a}
\Ra
 e^{   i {g} {\hat S}_{\rm R} }
{\cal D}_{\rm R}^{\dot \a \a}
 e^{ -  i {g} {\hat {\ov S}}_{\rm R} }
\ee
The invariant action is now:
\be
{\cal A} =
-\int d^4 x \; d^4 \q \; 
\lt \{
 {\hat {\ov \w}}_{{\rm L} \a}^{  T}
e^{-{g}{\hat V}_{\rm L}}  
{\cal D}_{\rm L}^{\dot \a \a}
 {\hat \w}_{{\rm L}\dot \a}
\rt \}\eb-\int d^4 x \; d^4 \q \; 
\lt \{
 {\hat {\ov \w}}_{{\rm R} \a}^{  T}
e^{-{g}{\hat V}_{\rm R}}  
{\cal D}_{\rm R}^{\dot \a \a}
 {\hat \w}_{{\rm R}\dot \a}
\rt \}\eb
-\int d^4 x \; d^2 \q \; 
\lt ( m^2 {\hat \w}_{{\rm L} \dot \b}^T 
 M {\hat \w}_{\rm R}^{ \dot \b}
\rt )
-\int d^4 x \; d^2 \ov \q \; 
\lt ( m^2 {\hat {\ov \w}}_{{\rm L}   \b}^T 
 {\ov M} {\hat {\ov \w}}_{\rm R}^{   \b}
\rt )
\ee
This is still invariant under supersymmetry and under gauge symmetry, provided that the mass terms are invariant under the gauge group.  In detail these terms have the form:
\be
{\hat \w}_{{\rm L} \dot \b}^T 
 M {\hat \w}_{\rm R}^{ \dot \b}
\equiv 
{\hat \w}_{{\rm L} \dot \b}^i 
 M_{ij} {\hat \w}_{\rm R}^{j \dot \b}
\ee
and we require that $M$ is an invariant numerical tensor of the gauge group just as for the  chiral scalar superfield:
\be
M_{ij} t^{aj}_{{\rm R}\;\;k}
+
t^{aj}_{{\rm L}\;\;i} M_{jk} =0
\ee

\subsection{ Action and Mass term for two Chiral Dotted Spinor Superfields $ {\hat A}_{{\rm L} \dot \a \dot \b}, {\hat A}_{{\rm R}\dot \a \dot \b }$  with two spinor indices  }

\la{doubledotsuperaction}

Now for the chiral doubledotsuper superfield we get:
\be
{\cal A} =\eb
\int d^4 x \; d^4 \q \; 
\lt (
 {\hat \A}_{{\rm L} \a \b}^T  e^{-{g} 
{\hat V_{\rm L}}}
 {\cal D}_{\rm L}^{\dot \a \a}
 {\cal D}_{\rm L}^{\dot \b \b}
 {\hat A}_{{\rm L} \dot \a\dot \b} 
 \rt )
\eb
+
\int d^4 x \; d^4 \q \; 
\lt (
 {\hat \A}_{{\rm R} \a \b}^T  e^{-{g} 
{\hat V_{\rm R}}}
 {\cal D}_{\rm R}^{\dot \a \a}
 {\cal D}_{\rm R}^{\dot \b \b}
 {\hat A}_{{\rm R} \dot \a\dot \b} 
 \rt )
\eb
\eb
+ \int d^4 x \; d^2 \q \; 
\lt (
m^3    {\hat A}_{{\rm L}}^{ \dot \a\dot \b T} M {\hat A}_{{\rm R} \dot \a\dot \b} 
 \rt )
+ \int d^4 x \; d^2 \ov \q \; 
\lt (
m^3    {\hat \A}_{{\rm L}}^{ \a  \b T}
\ov M   {\hat \A}_{{\rm R}   \a \b} 
 \rt )
\ee
and this is invariant assuming that the mass matrix is invariant as above for the single dotsuper, under the transformations: 
\be
{\hat A}_{{\rm L} \dot \a\dot \b} \Ra 
e^{i{g}  {\hat S_{\rm L}}} 
{\hat A}_{{\rm L}\dot \a\dot \b}
\ee
\be
{\hat A}_{{\rm R} \dot \a\dot \b} \Ra 
e^{i{g}  {\hat S_{\rm R}}} 
{\hat A}_{{\rm R}\dot \a\dot \b}
\ee

\subsection{ Action and Mass term for two Chiral Dotted Spinor Superfields $ {\hat \w}_{{\rm L}\dot \a \dot \b\dot \g }, {\hat \w}_{{\rm R} \dot \a \dot \b\dot \g }$  with three spinor indices  }

\la{tripledotsuperaction}

Similarly, for the chiral tripledotsuper superfield we get:

\be
{\cal A} =\eb
\int d^4 x \; d^4 \q \; 
\lt (
 {\hat {\ov \w}}_{{\rm L} \a \b \g }^T 
 e^{-{g} {\hat V}_{\rm L}} 
 {\cal D}_{\rm L}^{\dot \a \a}
 {\cal D}_{\rm L}^{\dot \b \b}
 {\cal D}_{\rm L}^{\dot \g \g}
 {\hat \w}_{{\rm L} \dot \a\dot \b\dot \g} 
 \rt )\eb
\int d^4 x \; d^4 \q \; 
\lt (
 {\hat {\ov \w}}_{{\rm R} \a \b \g }^T 
 e^{-{g} {\hat V}_{\rm R}}  
 {\cal D}_{\rm R}^{\dot \a \a}
 {\cal D}_{\rm R}^{\dot \b \b}
 {\cal D}_{\rm R}^{\dot \g \g}
 {\hat \w}_{{\rm R} \dot \a\dot \b\dot \g} 
 \rt )
\eb
+ \int d^4 x \; d^2 \q \; 
\lt (
m^4    {\hat \w}_{{\rm L}}^{ \dot \a\dot \b \dot \g T} M 
{\hat \w}_{{\rm R} \dot \a\dot \b \dot \g} 
 \rt )
+ \int d^4 x \; d^2 \ov \q \; 
\lt (
m^4    {\hat {\ov \w}}_{{\rm L}}^{ \a  \b \g T}\ov  M {\hat {\ov \w}}_{{\rm R}   \a \b \g} 
 \rt )
\ee
where
\be
{\hat \w}_{{\rm L} \dot \a\dot \b\dot \g} \Ra 
e^{i{g}  {\hat S}_{\rm L}} 
{\hat \w}_{{\rm L}\dot \a\dot \b\dot \g}
\ee
\be
{\hat {\ov \w}}_{{\rm R} \a \b\g} \Ra 
e^{ i{g}  {\hat {\ov S}}_{\rm R}} 
{\hat {\ov \w}}_{{\rm R} \a \b\g} 
\ee
and this is invariant under the gauge transformations and under supersymmetry transformations, 
assuming that the mass matrices are invariant as usual.

\section{Equation of Motion of Free Theory }

\la{doubledoteq}

\subsection{Summary of free equations of Motion for the doubledotsuper}

When the gauge coupling is taken to zero the gauge invariant derivative, acting on a chiral superfield, reduces to simply:
\be
{\cal D}_{\rm L}^{\dot \a \a}
\Ra 
\pa^{\dot \a \a}
\ee

So for the doubledotsuper we get:
\be
{\cal A}_{\rm Free} =\eb
\fr{1}{2}
\int d^4 x \; d^4 \q \; 
\lt (
 {\hat \A}_{{\rm L} \g \d }^T  \pa^{\dot \a \d}
 \pa^{\dot \b \g}   {\hat A}_{{\rm L} \dot \a\dot \b} 
 \rt )\eb
+\fr{1}{2}
\int d^4 x \; d^4 \q \; 
\lt (
 {\hat \A}_{{\rm R} \g \d }^T  \pa^{\dot \a \d}
 \pa^{\dot \b \g}   {\hat A}_{{\rm R} \dot \a\dot \b} 
 \rt )
\eb
+ \int d^4 x \; d^2 \q \; 
\lt (
m^3    {\hat A}_{{\rm L}}^{ \dot \a\dot \b T} M {\hat A}_{{\rm R} \dot \a\dot \b} 
 \rt )
+ \int d^4 x \; d^2 \ov \q \; 
\lt (
m^3    {\hat \A}_{{\rm L}}^{ \a  \b T} 
\ov M  {\hat \A}_{{\rm R}   \a \b} 
 \rt )
\ee
and here are two of the  equations of motion for this free action:
\be
\fr{\d {\cal A}}{\d {\hat \A}_{{\rm L} \g \d }^T } = 
\fr{1}{2}
D^2 \pa^{\dot \a \d}
 \pa^{\dot \b \g}   {\hat A}_{{\rm L} \dot \a\dot \b} 
+   
m^3  \ov M    {\hat \A}_{{\rm R}}^{   \g \d} 
=0
\la{eqwrwre1}
\ee
\be
\fr{\d {\cal A}}{\d \hat A_{\rm R}^{ \dot \e \dot \z T } }= 
\fr{1}{2}
\ov D^2 \pa_{\dot \e \d}
 \pa_{\dot \z \g}   {\hat \A}_{\rm R}^{ \g \d} 
+   
m^3  M^T    {\hat A}_{{\rm L}    \dot \e \dot \z } 
=0
\la{eqwrwre2}
\ee
Acting on (\ref{eqwrwre1}) with $\ov D^2 \pa_{\dot \e \d}
 \pa_{\dot \z \g} $ yields

\be
\ov D^2 \pa_{\dot \e \d}
 \pa_{\dot \z \g}\fr{\d {\cal A}}{\d {\hat \A}_{{\rm L} \g \d }^T } = 
\fr{1}{2}
\ov D^2 \pa_{\dot \e \d}
 \pa_{\dot \z \g}D^2 \pa^{\dot \a \d}
 \pa^{\dot \b \g}   {\hat A}_{{\rm L} \dot \a\dot \b} 
+   
m^3   \ov D^2 \pa_{\dot \e \d}
 \pa_{\dot \z \g} \oM  {\hat \A}_{{\rm R}}^{   \g \d} 
=0
\la{eqwrwre1mod}
\ee
and multiplying (\ref{eqwrwre2}) by $2m^3 \ov M$ yields
\be
m^3 \ov M\fr{\d {\cal A}}{\d \hat A_{\rm R}^{ \dot \e \dot \z T } }=  
\ov D^2 \pa_{\dot \e \d}
 \pa_{\dot \z \g}  m^3 \ov M {\hat \A}_{\rm R}^{ \g \d} 
+ 2  
m^6  \ov M  M^T    {\hat A}_{{\rm L}    \dot \e \dot \z } 
=0
\ee
and eliminating the common term yields
\be
\fr{1}{2}
\ov D^2 \pa_{\dot \e \d}
 \pa_{\dot \z \g}D^2 \pa^{\dot \a \d}
 \pa^{\dot \b \g}   {\hat A}_{{\rm L} \dot \a\dot \b} 
-   
2 m^6 \oM  M^T {\hat A}_{{\rm L} \dot \e\dot \z} 
=0
\la{eqwrwre1mod2}
\ee
and then using (\ref{eqwrwre2}) and
\be
[\ov D^2, D^2 ]
=- 4  \D
+ 4 D^{ \a} \ov D^{\dot \b}  \pa_{\a \dot \b} 
\ee
we get
\be
-2 ( \D^3    
+ m^6 \oM  M^T) {\hat A}_{{\rm L} \dot \e\dot \z} 
=0
\ee
The matrix $ m^6 \oM M^T$ is a hermitian positive semidefinite matrix, and it can be diagonalized to a set of positive eigenvalues $w$  times $m^3$.  So this equation reduces to a number of copies of the following:
\be
\lt (
  \D^3
+   w^3
m^6  \rt )
   {\hat A}_{{\rm L}    \dot \e \dot \z } 
\ee
which is equivalent to
\be 
\lt (
\D
+ w 
m^2  \rt )
\lt (
\D^2
- w m^2  \D +  w^2 m^4 \rt )
   {\hat A}_{{\rm L}    \dot \e \dot \z } 
=0
\la{eqwrwre1modnow}
\ee
Now this is interesting.  The factor $\lt (
\D
+ w 
m^2  \rt )$
is a normal factor for an equation of motion with mass 
$\sqrt{w} m$ in the present metric, as shown in \ci{cybersusyI}.  The other factor  
$\lt (
\D^2
- w m^2  \D +w^2 m^4 \rt )$ has no mass pole for physical values of the momentum. 

In fact we can put 
\be
X = \fr{\D}{ m^2} 
\ee
and then the above quadratic equation 
\be
\D^2
- w m^2  \D +w^2 m^4 =0 \Ra  
X^2 - w X + w^2 =0
\ee
has solutions
\be
X = \fr{-w + \sqrt{ - 3 w^2 }}{2}
,X = \fr{-w - \sqrt{ - 3 w^2 }}{2}
\ee
and clearly all its solutions are complex for any positive value of $w$. 

So for each eigenvalue of the matrix $m^6 \oM M^T$ there is one and only one mass, despite the fact that this equation of motion has $\D^3$ terms in it.

\subsection{Summary of free equations of Motion for Dotsupers  for the simplest cases, as reviewed above}

The equation of motion for the chiral scalar superfield, using the present notation,  was discussed in 
\ci{cybersusyI}, where we saw the equation
\be 
\lt (
\D
+  
w m^2  \rt )
   {\hat A}_{{\rm L}  } 
=0
\la{eqwrwreonescalar}
\ee

The equation of motion for the dotsuper was also discussed , using the present notation,  in 
\ci{cybersusyI}, where we saw the equation
\be 
\lt (
\D^2 -
w^2 m^4  \rt )
   {\hat \w}_{{\rm L}    \dot \e  } 
=
\lt (
\D
+  
w m^2  \rt )
\lt (
\D
- w m^2   \rt )
   {\hat \w}_{{\rm L}    \dot \e  } 
=0
\la{eqwrwreonedot}
\ee
This was the first example where the equation of motion is higher order in $\D$.  Here, although the solution of the extra term is not complex, it is positive, and no physical set of momenta has a zero for the expression $\lt (
\D
- w m^2   \rt )$.

As we saw in subsection \ref{doubledoteq}, for the doubledotsuper we get:
\be
\lt (
  \D^3
+   w^3
m^6  \rt )
   {\hat A}_{{\rm L}    \dot \e \dot \z } 
\eb
=
\lt (
\D
+ w 
m^2  \rt )
\lt (
\D^2
- w m^2  \D +  w^2 m^4 \rt )
   {\hat A}_{{\rm L}    \dot \e \dot \z } 
=0
\la{eqwrwre1modnow2}
\ee

  For the chiral tripledotsuper superfield we get
\be 
\lt (
\D^4
-  
m^8 w^4 \rt ){\hat \w}_{{\rm L}    \dot \e  \dot \z  \dot \h  } 
\eb
=\lt (
\D
+  
w m^2  \rt )
\lt (
\D^3
- \D^2
w m^2 +
\D w^2 m^4 
- w^3 m^6   \rt )
   {\hat \w}_{{\rm L}    \dot \e  \dot \z  \dot \h  } 
=0
\la{eqwrwrethreedot}
\ee
For this equation to behave similarly to the other two, the  cubic equation 
$\lt (
X^3
- X^2
w  +
X w^2  
- w^3    \rt )
=0$ needs to have no real negative solutions $X$, and indeed this is true. 
The solutions of the cubic equation
\be
\lt (
X^3
- X^2
w  +
X w^2  
- w^3    \rt )
=0
\ee
are
:
\be
 \{\{X\to -i w\},\{X\to i w\},\{X\to w\}\}
\ee
None of these are possible masses, so the threedotsuper behaves like the others.  

For the chiral fourdotsuper superfield we get the
 equation:
\be 
\lt (
\D^5
+  
m^{10} w^5 \rt )  {\hat \w}_{{\rm L}    \dot \e  \dot \z  \dot \h  \dot \z  } 
\eb
=\lt (
\D
+  
w m^2  \rt )
\lt (
\D^4
- \D^3
w m^2 
\ebp
+
\D^2 w^2 m^4 
-
\D w^3 m^6 
+ w^4 m^8   \rt )
   {\hat \w}_{{\rm L}    \dot \e  \dot \z  \dot \h  \dot \z  } 
=0
\la{eqwrwrethreedot2}
\ee
and this gives rise to the quartic equation
\be
\lt (
X^4
- X^3 w +
X^2 w^2  
- X w^3  
+
w^4  \rt )
=0
\ee
with solutions:
\be
\left\{\left\{X\to \frac{1}{4} \left(1+\sqrt{5}-i \sqrt{10-2
   \sqrt{5}}\right) w\right\},
\ebp
\left\{X\to \frac{1}{4}
   \left(1+\sqrt{5}+i \sqrt{10-2 \sqrt{5}}\right) w\right\},
\ebp
\left\{X\to
   \frac{1}{4} \left(-\sqrt{5} w+w-\sqrt{2 \left(5+\sqrt{5}\right)}
   \sqrt{-w^2}\right)\right\},
\ebp
\left\{X\to \frac{1}{4} \left(-\sqrt{5}
   w+w+\sqrt{2 \left(5+\sqrt{5}\right)}
   \sqrt{-w^2}\right)\right\}\right\}
\ee
None of these solutions are negative and real. So the fourdotsuper behaves like the others.  Only one solution corresponds to a mass.

\subsection{Summary of free equations of Motion for Dotsupers for the general case, with a conjecture}

The generalization is clear. The bosonic dotsupers   have actions that are the generalization of the action in subsection \ref{doubledotsuperaction}, and the fermionic  dotsupers  have actions that are the generalization of the actions in subsections \ref{singledotsuperaction}
 and \ref{tripledotsuperaction}.

 The general  free equation of motion for fermionic dotsupers is of the form
\be 
\lt (
\D^{2n+2}
-  
m^{4n+4} w^{2n+2} \rt ){\hat \w}_{{\rm L}  \dot \a_1 \cdots \dot \a_{2n+1}   } , n =0, 1,2,\cdots
\ee
and for any  $ n = 0,1,2,\cdots$ this can be written
 \[
\lt (
\D
+  
w m^2  \rt )
\]
\be
\lt (
\D^{2n+1}
- \D^{2n}
w m^2 + \cdots
- w^{2n+1} m^{4n+2}   \rt )
   {\hat \w}_{{\rm L}    \dot \a_1 \cdots \dot \a_{2n+1}  } 
=0
\la{generalfermidoteq}
\ee
The general  free equation of motion for bosonic dotsupers is of the form
\be 
\lt (
\D^{2n+1}
+  
m^{4n+2} w^{2n+1} \rt ){\hat A}_{{\rm L}  \dot \a_1 \cdots \dot \a_{2n}   } , n = 0,1,2 \cdots 
\ee
and for $n = 1,2,\cdots $ this can be written:
\[
\lt (
\D
+  
w m^2  \rt )
\]
\be
\lt (
\D^{2n}
- \D^{2n-1}
w m^2 + \cdots
+ w^{2n} m^{4n}   \rt )
   {\hat A}_{{\rm L}    \dot \a_1 \cdots \dot \a_{2n}  } 
=0
\la{generalbosedoteq}
\ee

This yields the following polynomial equations for the roots of the second factors above:
\be
P_{\rm Fermi,2n+1} = \lt (
X^{2n+1}
- X^{2n}
w  + \cdots
- w^{2n+1}    \rt )
=0, n = 0,1,2\cdots
\ee
\be
P_{\rm Bose,2n} = \lt (
X^{2n}
- X^{2n-1}
w  + \cdots
+ w^{2n}    \rt )
=0, n = 1,2\cdots
\ee
and the conjecture is that these polynomial equations $P_{\rm Fermi,2n+1} =0$ and $P_{\rm Bose,2n} =0$ do not have any negative real roots for any of the indicated integer value of n. 
We have proved the conjecture for the Fermi case for n=0,1 and for the Bose case for n=0,1,2.

Assuming the conjecture is true in general would imply that for all cases of dotsupers with any number of indices, the  equation of motion has a normal factor $\lt (
\D
+  
w m^2  \rt )$ with a mass  at $\sqrt{w} m $,  times a factor with no further mass solution for physical values of the momenta.    

If Cybersusy makes any sense, these  should all be viable equations of motion, with viable propagators, and no serious violation of any physical principles.  Whether that is true or not remains unclear. If it is true, it is an interesting fact in its own right.  For some comments on the history of equations of motion for higher spin fields, see 
\ci{weinberg1}.

\section{General Remarks} 

\subsection{ 
The vector boson fields in the dotsupers are not part of gauge boson multiplets, although they may be coupled to gauge boson multiplets}

Note that there are higher spin fields embedded in these superfields.  For example ${\hat \w}_{{\rm L}    \dot \e   } $ contains a vector boson field and two spinors and  a scalar.  The vector boson is not a gauge boson.  Does this make physical sense?  In Cybersusy, after gauge and supersymmetry breaking,  this vector boson carries lepton number, which is conserved.  A gauged vector boson cannot carry any conserved quantum number, because if it did, then its gauge transformation parameter S in 
\be
V_{\a \dot \b} \ra 
\pa_{\a \dot \b} S + \cdots 
\ee
would carry the same quantum number, and then $e^{iS}$ would be nonsense, since it would carry different quantum numbers in its expansion terms.  But a vector boson inside a chiral dotted spinor superfield like ${\hat \w}_{{\rm L}    \dot \e   } $ can apparently carry a conserved quantum number, because it has no gauge transformation. Moreover, the quantum number remains conserved after supersymmetry breaking.    

It is important for Cybersusy that things work this way, because Cybersusy needs to account  for the lack of observed supersymmetry in the baryons and  the leptons, and baryon number and lepton number are conserved, and there are plenty of vector bosons with baryon number and lepton number in Cybersusy.

\subsection{Coupling of Dotsupers to Composite Superfields}

\la{compdotsupers} 

A curious feature emerges early on.  Suppose that we start with a model with chiral scalar superfields and gauged vector superfields.  Suppose that we do not start with any chiral dotted spinor superfields. 

As explained in \ci{cybersusyI}\ci{cybersusyII}\ci{cybersusyIII}\ci{cybersusyIV}\ci{cybersusyV}\ci{susy09}, it is natural to put together composite chiral dotted spinor superfields in such a theory.  These are made from the components of the chiral scalar superfields, combined with the Zinn-Justin sources for the variations of these components. 
 They act like chiral dotted spinor superfields under the action of the BRS transformations.  So we called them chiral dotted spinor pseudosuperfields, or dotpseudos for short.

 The simplest ones have forms like
\be
{\hat \w}_{{\rm Pseudo}\;\dot \a} = f^{i}_{j} {\hat {\ov \y}}_{{\rm Fund}\; i \dot \a}
 {\hat A}_{\rm Fund}^{j} 
\la{rightstuff}
\ee
where
\be
{\hat A}_{\rm Fund}^{i}(x) = A^i(y) +
\q^{\a} 
\y^i_{\a } (y) 
+ \fr{1}{2} \q^{\g} \q_{\g} 
G^i(x)
\ee
and 
\be
{\hat {\ov \y}}_{{\rm Fund}\; i \dot \a}(x) = \ov \y_{i \dot \a }(y) 
+
\q^{\b} 
\lt [
\pa_{\b \dot \a} \A_i(y) 
+ \ov C_{\dot \a} Y_{i  \b}(y) \rt ]
- \fr{1}{2} \q^{\g} \q_{\g} 
\G_{i}(x)  
\ov C_{\dot \a} 
\ee

The transformation induced by $\d_{\rm BRS} $ is summarized by the following equation:
$
\d_{\rm BRS}  {\hat A}_{\rm Fund}^{i}(x)=   \d_{\rm SS} {\hat A}_{\rm Fund}^{i}(x) $ where the superspace operator is $\d_{\rm SS} =
C^{\a} Q_{\a} 
+ \ov C^{\dot \a} \ov Q_{\dot \a} 
$.  This relation means that the effect of  $\d_{\rm BRS}$ on this particular combination is the same as the effect of the superspace operator $\d_{\rm SS}$.  The supertranslations are: 
$
 Q_{\a} = \fr{\pa}{\pa \q^{\a}} -  \fr{1}{2} \pa_{\a \dot \b} \oq^{\dot \b} $ and $  \oQ_{\dot \a} = \fr{\pa}{\pa \oq^{\dot \a}} -  \fr{1}{2} \pa_{\b \dot \a} \q^{\b}
$.

Would it be sensible to couple independent chiral dotted spinor superfields to these composite chiral dotted spinor superfields in such a theory?  Does the situation call for the addition of 

\be
\int d^4 x\; d^2 \q \; {\hat \w}_{{\rm Super}}^{\dot \a} 
{\hat \w}_{{\rm Pseudo}\;\dot \a} 
\ee
to the action?

The answer here is clearly {\bf NO}.  The reason is that the transformation of ${\hat \w}_{{\rm Pseudo}\;\dot \a} $ as a chiral superfield uses the transformations of the Zinn Fields which arise after the auxiliaries are integrated.  However one must not integrate these before one adds all the fields in the theory.  If there are fundamental dotsupers in the theory, they 
must be coupled before the auxiliaries are integrated, not after. 

This is why it is natural to examine the effective action for the cybersusy theory.  It makes no sense to couple the resulting composite dotted pseudosuperfields to external sources or to fundamental dotsupers.

On the other hand it is also tempting to examine the theory with fundamental dotsupers coupled to composite dotsupers.  These have the form
\be
\int d^4 x\; d^2 \q \; {\hat \w}_{{\rm Super}}^{\dot \a} 
{\hat \w}_{{\rm Comp}\;\dot \a} 
\ee
 where
\be
{\hat \w}_{{\rm Comp}\;\dot \a} 
= 
\ot^{ij} \ov D^2 \lt \{ {\hat \A}_i \ov D_{\dot \a}  {\hat \A}_j \rt \} 
\ee
or 
\be
{\hat \w}_{{\rm Comp}\;\dot \a} 
= 
\ot^{ijk} \ov D^2 \lt \{ {\hat \A}_i \ov D_{\dot \a}  {\hat \A}_j {\hat \A}_k \rt \} 
\ee
etc.  It is not clear what the relation between such theories and cybersusy is.  That requires investigation.  Note that one can assume that these tensors give zero when symmetrized over all their indices:
\be
\ot^{(ij)} =0
\ee
and
\be
\ot^{(ijk)} =0
\ee

This arises because, for example,  the symmetric part $\ot^{(ijk)}$
gives rise to
\be
\ot^{(ijk)} \lt \{ {\hat \A}_i \ov D_{\dot \a} ( {\hat \A}_j {\hat \A}_k ) \rt \} 
= 
2 \ot^{(ijk)} \lt \{ {\hat \A}_i   {\hat \A}_j \ov D_{\dot \a} {\hat \A}_k  \rt \} 
= 
\fr{2}{3}
\ot^{(ijk)}   \ov D_{\dot \a} \lt \{ {\hat \A}_i   {\hat \A}_j {\hat \A}_k \rt \} 
\ee
and then we get a triple chiral derivative, which is zero:
\be
{\hat \w}_{{\rm Comp}\;\dot \a} 
= 
\ov D^2   \ov D_{\dot \a} \fr{2}{3}
\ot^{(ijk)} \lt \{ {\hat \A}_i   {\hat \A}_j {\hat \A}_k \rt \} 
=0
\ee

These bear a close relation to the constraints that one finds in Cybersusy.  This requires further investigation.  How does one relate Cybersusy to actions of this kind?

\subsection{Generalized Cybersusy Algebra and some Guesses}

\la{guesses}

The introductory papers for Cybersusy \ci{cybersusyI}\ci{cybersusyII}\ci{cybersusyIII}\ci{cybersusyIV}\ci{cybersusyV}
contain one repeated error, which was mentioned in \ci{susy09}. The problem is that  under weak SU(2) for the SSM, the leptons and quarks  are right handed chiral superfield singlets,   and left handed chiral superfield doublets. Combining this with the structure of the Higgs/Goldstone fields for the SSM, one can see that the right handed composite fields for the leptons have the form  
in equation (\ref{rightstuff}), and the left handed composite fields for the leptons have the form 
\be
{\hat \w}_{{\rm Pseudo}\;\dot \a} = f^{i}_{jk} {\hat {\ov \y}}_{{\rm Fund}\; i \dot \a}
 {\hat A}_{\rm Fund}^{j} 
 {\hat A}_{\rm Fund}^{k} 
\la{leftstuff}
\ee
For the specific case of that model one can vary the latter expression, after gauge symmetry breaking,  to  the following form:
\be
{\hat \w}_{{\rm Pseudo}\;\dot \a} =  {\hat {\ov \y}}_{{\rm Fund}\; i \dot \a}
 \lt ( f^{i}_{jk} (m v^j +  {\hat A}_{\rm Fund}^{j} )
(m v^k +  {\hat A}_{\rm Fund}^{k})  -m^2 f^{i} \rt )
\la{leftstuffnew}
\ee
and this form does not generate the term $ \d_{\rm L, Mix}$ which gives rise to left cybersusy. This problem can be expected to extend to the baryons and other particles in those papers.

Leaving the specific details of the SSM, and considering more general models of the same kind, we shall now discuss the general issues relating to left and right cybersusy. 

In general one might expect to get the following algebra:

\be
\d_{\rm L, Mix}  {\hat \w}^{i} _{{\rm L}(\dot \a_1 \cdots \dot \a_{2n+1})}
= f^{i}_{{\rm L}, n, j} \oC_{(\dot \a_1}  {\hat A}^{j}_{{\rm L}\dot \a_2 \cdots \dot \a_{2n+1})}
\ee

\be
\d_{\rm L, Mix} {\hat A}^{i}_{{\rm L}(\dot \a_1 \cdots \dot \a_{2n})}
= b^{i}_{{\rm L}, n,j} \oC_{(\dot \a_1}  {\hat \w}^{j}_{{\rm L} \dot \a_2 \cdots \dot \a_{2n})}
\ee

Nilpotence conditions on the coefficients arise from the following.  

Firstly:
\be
\d_{\rm L, Mix}^2 {\hat \w}^{i} _{{\rm L}(\dot \a_1 \cdots \dot \a_{2n+1})}
= f^{i}_{{\rm L}, n, j}  b^{j}_{{\rm L}, n,k} \oC_{(\dot \a_1}  \oC_{\dot \a_2}  {\hat \w}^{k}_{{\rm L} \dot \a_3 \cdots \dot \a_{2n+1})}
=0
\ee
This is satisfied if 
\be
f^{i}_{{\rm L}, n, j}  b^{j}_{{\rm L}, n,k}=0
\ee
Secondly:
\be
\d_{\rm L, Mix}^2 {\hat A}^{i} _{{\rm L}(\dot \a_1 \cdots \dot \a_{2n})}
= b^{i}_{{\rm L}, n, j}  f^{j}_{{\rm L}, n-1,k} \oC_{(\dot \a_1}  \oC_{\dot \a_2}  {\hat A}^{k}_{{\rm L}\dot \a_3 \cdots \dot \a_{2n})}
=0
\ee
This is satisfied if 
\be
b^{i}_{{\rm L}, n, j}  f^{j}_{{\rm L}, n-1,k}=0
\ee
Building up from the bottom, we get:
\be
f^{i}_{{\rm L}, 0, j}  b^{j}_{{\rm L}, 0,k}=0
\ee
\be
b^{i}_{{\rm L}, 1, j}  f^{j}_{{\rm L}, 0,k}=0
\ee
\be
f^{i}_{{\rm L}, 1, j}  b^{j}_{{\rm L}, 1,k}=0 
\ee
\be
b^{i}_{{\rm L}, 2, j}  f^{j}_{{\rm L}, 1,k}=0
\ee
\be
\cdots 
\ee

This series can end at any point, with
\be
f^{i}_{{\rm L},n, j} =0 \; {\rm for} \; n \geq N_{\rm f L}
\ee
and 
\be
b^{i}_{{\rm L}, N_{\rm L}, j} =0 \; {\rm for} \; n \geq N_{\rm bL}
\ee
for any integers $N_{\rm fL}\geq 0$ and $N_{\rm bL}\geq 0$.

This algebra can arise independently for the left dotsupers or the right dotsupers or both.  In \ci{cybersusyI}\ci{cybersusyII}\ci{cybersusyIII}\ci{cybersusyIV}\ci{cybersusyV}, we saw a number of examples of this situation for the baryons for example, where the indices above turn out to be combinations of multiple weak isospin, and quark colour indices. 

Then it is natural to make the following guesses, which are a simple generalization of the results for dotspinors with one spinor index discussed in \ci{cybersusyI}\ci{cybersusyII}\ci{cybersusyIII}\ci{cybersusyIV}\ci{cybersusyV}\ci{susy09}:
\ben
\item
For any of these left algebras there is a left kinetic action 
such that 
\be
  \d_{\rm CYBERSUSY}  {\cal A}_{\rm L, Kinetic} = 
\lt ( \d_{\rm SUSY} + \d_{\rm L, Mix} \rt \} {\cal A}_{\rm L, Kinetic} = 0
\ee
\item
For the left algebra we find that the mass terms, which are invariant under SUSY: 
\be
\d_{\rm  SUSY} {\cal A}_{\rm Mass} =0
\ee
are not invariant under $\d_{\rm L, Mix}$:
\be
\d_{\rm L, Mix} {\cal A}_{\rm Mass} = {\cal A}_{\rm L, Anomaly}
\ee
 
\item
Furthermore, although the terms ${\cal A}_{\rm L, Anomaly}$ satisfy:
\be
\d_{\rm  SUSY}  {\cal A}_{\rm L, Anomaly}=0
\ee
there is no local expression ${\cal A}_{\rm L, Counterterm}$ such that
\be
{\cal A}_{\rm L, Anomaly} = \d_{\rm  SUSY}  {\cal A}_{\rm L, Counterterm}
\ee
\item
This whole system can be repeated for the right algebra with 
${\rm L} \Ra {\rm R}$.  The left and right algebras may be quite different in detail.  For the SSM they are in fact very different, and the left algebra appears to be trivial, while the right algebra is non-trivial. 
\item
Any such system generates a spectrum of SUSY breaking when we include the mass terms.  Different versions of the SSM generate different spectra. 
\item
 The simplest SSM, for the leptons, generates only the right Mixing, and it is not consistent with experiment--a paper on this topic is in preparation.
\item
If we could find a version of the SSM that generates both left and right mixing for the leptons, it might be consistent with experimental results for the leptons, as far as they are presently known. 

\een

\subsection{Conclusion}

In this paper we have exhibited actions for these higher spin chiral dotted  spinor superfields.  The actions are invariant under supersymmetry and gauge invariance.  We have also seen that the free theories do have a conventional interpretation in terms of particles and masses, even though the equations of motion contain powers of the D'Alembertian operator $\D$.

 A natural future step  is to finish the classification of the composite superfields in the SSM, and show that the Cybersusy mechanism applies to all or most of the observable particles.  This work is well under way, but it is complicated and lengthy.  It appears that most or all observed particles participate in Cybersusy.  Later these dotted chiral spinor superfields need to be analyzed for the supersymmetry breaking spectrum by writing down and diagonalizing the quadratic actions that result from Cybersusy. 

However this program is not very important unless one can find a variation of the SSM which is consistent with experiment for the leptons.  This probably  means that the model must have left and right operators 
$\d_{\rm L, Mix}$ and $\d_{\rm R, Mix}$ for the leptons.

Another line of inquiry is to determine how to make  Cybersusy a unitary theory, in spite of the presence of supersymmetry anomalies.  This is connected with the issue of the calculation of  decay amplitudes for the particles in the broken supersymmetry multiplets.  The coupling to the gauge theory, and the coupling to composite dotspinors made from chiral scalar superfields,  as set out in this paper, are preliminary steps for the examination of these problems.

\tableofcontents

\end{document}